\documentclass[twocolumn,trackchanges]{aastex7}

\begin{document}

\title{New Molecular Dynamics Methods for Simulating Neutron Star Crusts with Superfluid Vortices}

\author[orcid=0009-0006-2122-5606]{M. E. Caplan}
\affiliation{Department of Physics, Illinois State University, IL, USA}
\email[show]{mecapl1@ilstu.edu}  

\author[orcid=0009-0007-3110-9198]{N. T. Smith}
\affiliation{Department of Physics, Illinois State University, IL, USA}
\email{} 

\begin{abstract}
Superfluid vortices in neutron star crusts are thought to be pinned to the lattice of nuclei in the crust. The unpinning of superfluid vortices in spin glitches therefore motivates us to study the vortex-crust interaction explicitly with molecular dynamics. In this work, we present a new molecular dynamics methods to characterize the response of the crust to a rigid vortex. When vortex pinning forces and nearest neighbor Coulomb forces are comparable, we observe a qualitatively new phenomena of lattice entrainment with implications for the elastic evolution of the crust.
\end{abstract}

\keywords{\uat{Neutron stars}{1108} --- \uat{Plasma physics}{2089} --- \uat{Computational methods}{1965} --- \uat{Stellar interiors}{1606}  --- \uat{N-body simulations}{1083}}


\section{Introduction}

Neutron star spin glitches are now thought to be due to the rapid unpinning and repinning of an array of superfluid vortices in the crust, allowing the crust to briefly couple to the core and exchange angular momentum. The crust-vortex interaction, or more specifically the nucleus-vortex interaction, is therefore of utmost importance in understanding spin glitches and associated outbursts from magnetized neutron stars. 

Until now, studies of vortex pinning and unpinning have used static idealized crusts when simulating dynamical vortices using the equation of motion from \cite{Schwarz1977,Schwarz1978}, such as in \cite{Link2022}. The microphysical response of the crust to moving vortices is therefore unresolved and remains largely unstudied. Forward progress would be most readily achieved by coupling molecular dynamics codes that simulate the crustal lattice explicitly to a dynamical simulation of a vortex. As a step toward that, we present here new molecular dynamics capabilities for simulating crusts with idealized vortices.

\section{Methods}

Our molecular dynamics formalism follows from our recent work, \cite{Caplan2024}, and the works cited therein. 

Nuclei are treated as classical point particles of charge $eZ_i$  interacting via a screened Coulomb potential

\begin{equation}
    V(r_{ij}) = \frac{ e^2 Z_i Z_j }{r_{ij}} e^{-r_{ij}/\lambda}
\end{equation}

\noindent with separation $r_{ij}$ and screening length $\lambda$ typically taken to be the Thomas-Fermi screening length. 

In addition to this conservative pair potential, we introduce an external classical potential to our simulations following the suggestion in \cite{Link2022}. In that work, the nucleus-vortex interaction is treated as a Gaussian well of energy $E_p$ and characteristic width $\sigma$, which we adapt for a vortex aligned with the $z$ axis of our simulation volume by writing

\begin{equation}\label{eq:V}
    V(x,y)= E_p \exp[-(\Delta x^2+ \Delta y^2)/2 \sigma^2]
\end{equation}

\noindent such that a particle at position $\mathbf{r} = (x,y,z)$ experiences a force determined by $\Delta x = x - x_w$ where $x_w$ is the dynamical position of the vortex. The potential introduces a new energy-scale ($E_p$), length-scale ($\sigma$), and time-scale (vortex velocity) to the system, and so one might expect interesting new dynamics to emerge in frustrated systems where these scales are comparable to the ion plasma scales of the lattice.

\begin{figure*}
\includegraphics[width=0.98\textwidth]{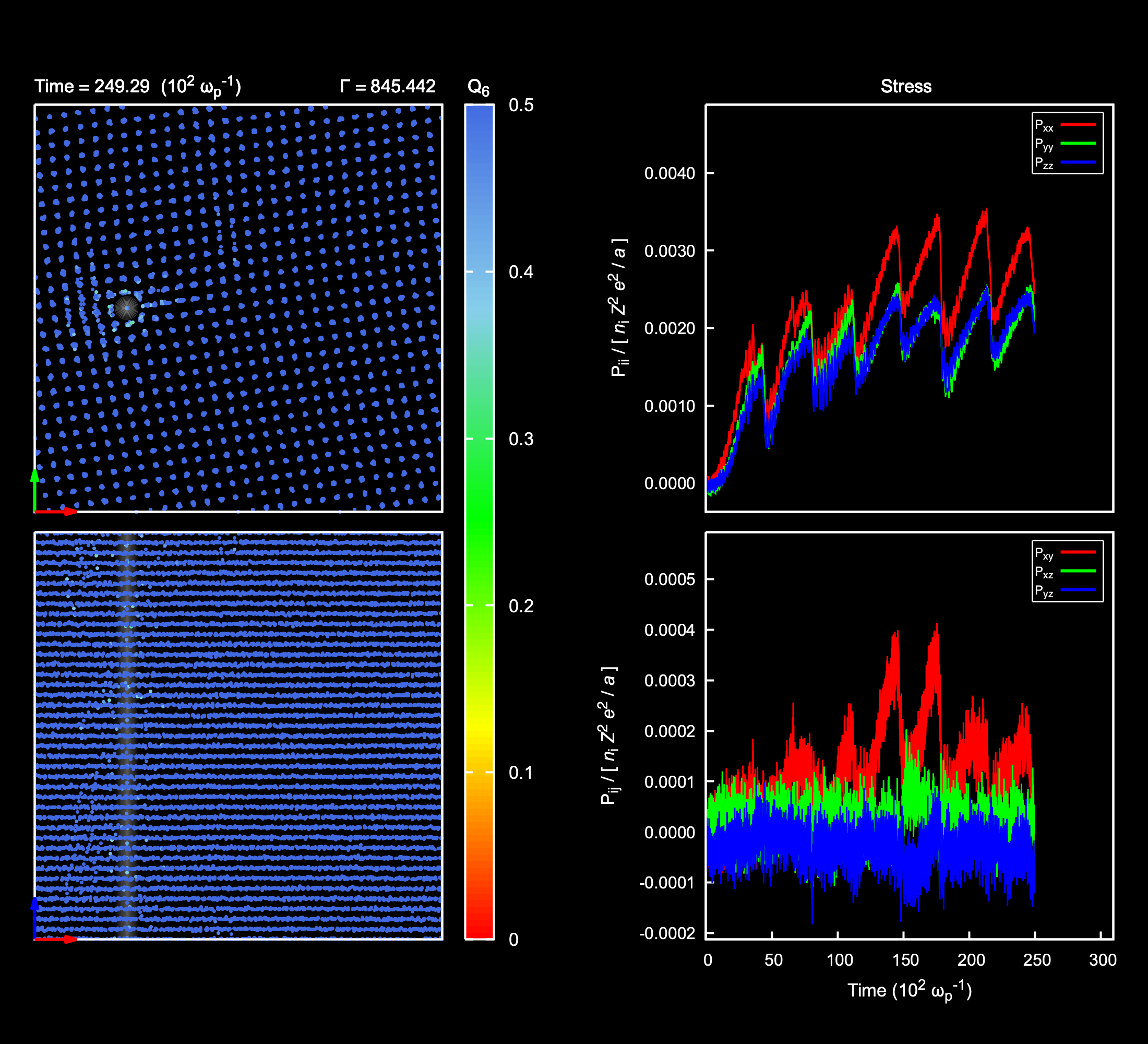}
\caption{\label{fig:fig1} {\bf Simulation animation of a displaced vortex.} Clockwise from bottom left: (1) Side view (xz plane) of our simulation volume with vortex (shaded grey column). (2) Top down view of our simulation (xy plane) along the vortex axis. Nuclei are colored by the bond order parameter $Q_6$, with blue (yellow-green) denoting ordered (disordered) nearest neighbors. (3) Diagonal elements of the stress tensor show tensile stress, which reaches a maxima just before breaking when the vortex has displaced by one lattice spacing. (4) Off-diagonal elements of the stress tensor show shear stresses. Animations available in the arXiv ancillary files. Mirrors: \href{https://i.imgur.com/YDOCCjH.mp4}{[1]}\href{https://i.imgur.com/QREYGkv.mp4}{[2]} }
\end{figure*}

\section{Simulation}

As a proof of concept, we simulate with parameters chosen so that the characteristic well force is ${E_p / \sigma \approx e^2 Z^2 / a_i^2 }$ to demonstrate this frustration. We use $Z=10$, $A=40$ nuclei at a number density $n_i=7.18 \times 10^{-5} \ \rm{fm}^{-3}$ for Wigner–Seitz radius ${a_i = 14.9255 \ {\rm fm}}$ (density $\rho = 2\times10^{13} \, \rm{g \, cm}^{-3}$ assuming a free neutron fraction of 0.75). Such low $Z$ would be unexpected in the inner crust of a cold catalyzed neutron star, but could be found in accreted neutron stars near pycnonuclear fusion densities \citep{fantina2018crustal}. 

The energy and length-scale of the vortex interaction are uncertain and likely vary with depth in the crust. In this work, we simulate with well parameters ${E = 4 \ {\rm MeV}}$ and ${\sigma = 5 \ {\rm fm}}$ at approximately ${T = 0.01 \ {\rm MeV} = 0.2 \, T_{\rm M}}$  (melting temperature $T_{\rm M}$, coupling ${\Gamma = e^2 Z^2 / a_i T = 845}$). 

The simulation in Fig. \ref{fig:fig1} is run for $680\,000$ timesteps using $dt = 25 \, \rm{fm/c}$ ($\omega_p^{-1}/21.5$). The lattice is rotated by $\arctan(4/5)$ in the xy-plane so that the vortex displacement is slightly misaligned with the lattice. An attractive vortex along the $z$ axis aligned with the bcc (100) plane is displaced in $-x$ producing a sawtooth stress response, with periodic breaks occurring when the well is displaced by approximately a lattice spacing.  
The vortex crosses one unit cell of the crystal approximately every $10^4$ plasma oscillations.

For the crystal (left), we show a top down view of the xy plane (top) and a side view of the xz plane (bottom), with the vortex potential highlighted in grey. The differential pressure tensor (right) gives the tensile stress (top) and shear stress (bottom). 

The vortex is initially aligned with a stack of 20 nuclei, acting as a trap potential. As the vortex drifts it carries that column of nuclei with it, such that the nearest neighbors to the vortex feel a repulsive Coulomb force from the nuclei bound to the vortex. Therefore, an attractive vortex is effectively repulsive at sufficiently strong pinning and sufficiently low temperatures that the nuclei cannot become easily become unbound from the vortex. 

As the well displaces to the left the lattice bows globally in response. Locally, the lattice deforms due to the displacement of the line of nuclei bound to the vortex potential resulting in large localized stresses that can be seen in the smaller bond order parameter $Q_6$ (colorbar). When the vortex is displaced by a lattice spacing, the nearest neighbors ahead of the vortex are compressed and displaced while the nuclei on the vortex leave a void behind them. The nuclei bound to the vortex are entrained and transported through the material, and when stresses are sufficiently large the `upstream' nuclei diffuse around them into the vacancy. While these diffusive hops are thermally activated, diffusion coefficients likely scale strongly with the lattice deformation, enabling in the rapid breaking \citep{Caplan2024}. 
Stresses are relieved by this diffusive event and the lattice is once again an ideal bcc crystal in equilibrium. 

Additional nuclei become bound to the vortex following the first few breaking events. This changes the nearest neighbor spacing of the nuclei bound to the vortex causing those nuclei to form a defect line. This mismatched spacing may lower the threshold for breaking in the subsequent steps, and is also a means of introducing vacancies into the surrounding crystal. One such defect can be seen just above the center of the box in the snapshot in Fig. \ref{fig:fig1}.

\section{Summary}

We argue that vortex pinning and unpinning may be highly frustrated, and when the pinning force is comparable to the nearest neighbor Coulomb force these systems enter the qualitatively new entrainment regime observed here. Lattice entrainment likely alters the unpinning threshold and lattice response in the neutron star crust at the depths where it is active, and should be fully characterized in future work. 

Even in this simple model with a rigid vortex, we expect there is a rich phase diagram for unpinning behavior that depends on the relative vortex and Coulomb forces as well as vortex radius. This behavior is likely glassy making it difficult to characterize analytically but straightforward to study with molecular dynamics. 
With these new capabilities in hand this should be relatively straightforward, and should efficiently identify interesting regimes to study with more complicated simulations in next generation codes that directly couple dynamic vortex solvers to molecular dynamics.

\begin{acknowledgments}
Financial support for this publication comes from Cottrell Scholar Award \#CS-CSA-2023-139 sponsored by Research Corporation for Science Advancement. This work was supported by a grant from the Simons Foundation (MP-SCMPS-00001470) to MC. This research was supported in part by the National Science Foundation under Grant No. NSF PHY-1748958.
\end{acknowledgments}

\end{document}